\documentclass[12pt]{article}
\usepackage[dvips]{graphicx}
\date{\today}
\newcommand{\CP}{
\begin{picture}(20,10)
\put(6.5,-3){\line(2,1){15}}
\put(0,0){$P_{CP}$}
\end{picture}}

\begin{document}
\begin{titlepage}
\renewcommand{\thefootnote}{\alph{footnote}}
\begin{flushright}
  {DPNU 00-24}\\
  {August 2000}
\end{flushright}
\vspace*{0.5cm}
\renewcommand{\thefootnote}{\fnsymbol{footnote}}
\setcounter{footnote}{-1}
{\begin{center}
{\large\bf Schematic Method for Estimation of CP Asymmetry \\ 
in Neutrino Oscillation}
\end{center}}
\renewcommand{\thefootnote}{\alph{footnote}}
\vspace*{.8cm}
{\begin{center} {\large{\sc
K.~Kimura\footnote[1]{\makebox[1.cm]{Email:}
kimukei@eken.phys.nagoya-u.ac.jp}
\vspace{0.2cm}  and
{\sc A.~Takamura\footnote[2]{\makebox[1.cm]{Email:}
takamura@eken.phys.nagoya-u.ac.jp}\footnotemark[3]}}}
\end{center}}
\vspace*{0cm}
{\it \begin{center}     \footnotemark[1]${}^,$\footnotemark[2]%
Department of Physics, Nagoya University \\
Nagoya 464-8602, Japan
\vskip .3cm
\footnotemark[3]%
Department of Mathematics, Toyota National Collage of Technology
 Eisei-cho 2-1, Toyota-shi, 471-8525, Japan
\end{center} }
\vspace*{0.5cm}
{\Large \bf \begin{center} Abstract \end{center}  }%
\hspace*{\parindent}
Within the framework of three generations of leptons
the schematical method for estimating CP asymmetry
in neutrino oscillations is considered.
We introduce a unitarity triangle corresponding to 
$\nu_e$-$\nu_{\mu}$ oscillation,
in addition, we show that it is convenient for the estimation
to define another new triangle.
CP asymmetry is determined
by the difference between the shapes of a
unitarity triangle and new triangle.
As results,
(i) we show that CP asymmetry becomes
maximal if the shapes of these two triangles are the same.
(ii) We can easily estimate $L/E$
which leads almost $100\%$ asymmetry using this.
(iii) In $\nu_e$-$\nu_{\mu}$ oscillation with
$\sin^2 2\theta_{13} \simeq 0.04$, 
we obtain about $90\%$ asymmetry for LMA MSW scenario
and $3\%$ asymmetry for SMA MSW scenario within
long baseline neutrino expriments 
to be realized in near future.
\end{titlepage}

\section{Introduction}

\hspace*{\parindent}
Recent atmospheric neutrino experiments
\cite{SKatm}
strongly suggest $\nu_{\mu}$-$\nu_{\tau}$ oscillations
with large mixing.
On the other hand, solar neutrino experiments
\cite{GALLEX}
also suggest $\nu_e$-$\nu_{\mu}$ or $\nu_e$-$\nu_{\tau}$
oscillations.
These results imply the participation of at least
three generations in the lepton sector.

In the framework of three generations, it is natural
to consider CP violation in neutrino oscillations.
The estimation of CP phase in the lepton sector,
if it exists, is very important
to construct the physics beyond the Standard Model.
As $1$-$3$ mixing and $1$-$2$ mass 
difference are small, it is difficult to observe 
CP violation effects at present experiments.
However, there are following favorable points to observe
CP violation in neutrino oscillations.
First,
various long baseline experiments with high 
intensity neutrino beams are planned in near future. 
There are mainly two kinds of experiments with 
high energy neutrino like neutrino factory experiments 
\cite{neutrino factory}, 
and with low energy neutrino like PRISM \cite{PRISM}.
Second, 
the appearence of large $2$-$3$ mixing 
discovered by Super-Kamiokande Collaboration 
provides the
possibilities of large CP violation in the lepton sector.
This is contrast to CP violation in the quark sector
with small $2$-$3$ mixing.
Considering of this point,
CP violation in neutrino oscillations
have been investigated by many authors
\cite{Sato, Tanimoto}.

In this work, we propose the schematical approach 
to estimate CP asymmetry
in neutrino oscillation defined by
\begin{eqnarray}
A_{CP}&=&\frac{\Delta \CP}
{\Delta P_{CP}}=
\frac{P(\nu_{\alpha} \to \nu_{\beta})
-P(\bar{\nu}_{\alpha} \to \bar{\nu}_{\beta})}
{P(\nu_{\alpha} \to \nu_{\beta})
+P(\bar{\nu}_{\alpha} \to \bar{\nu}_{\beta})},
\end{eqnarray}
where $\Delta P_{CP}$ is CP-conserving part and
$\Delta \CP$ is CP-violating part.
We introduce a unitarity triangle
in the lepton sector 
and another new triangle to estimate $A_{CP}$.
We explore the condition in which
$A_{CP}$ takes the value as large as possible
using the geometry of the two triangles.
From the condition, $L/E$ which leads the almost
$100\%$ asymmetry can be estimated easily,
where $L$ is the baseline length and
$E$ is the neutrino energy.
In the case of LMA and SMA MSW scenarios,
we derive such $L/E$ and estimate whether 
large $A_{CP}$ is realized in 
near future experiments.

\section{Neutrino Oscillations in Three Generation}

\hspace*{\parindent}
In this section we consider 
the neutrino oscillation in three generation 
and represent transition probability
from one flavor to another flavor
using the components of the Maki-Nakagawa-Sakata matrix 
(MNS matrix) \cite{MNS}.

In vacuum, flavor eigenstates
$\nu_{\alpha}\, (\alpha=e,\mu,\tau)$
are related to mass eigenstates $\nu_i(i=1,2,3)$,
which have the mass eigenvalues
$m_i$, by unitary transformation,
\begin{equation}
\nu_{\alpha}=U_{\alpha i}\nu_i,
\end{equation}
where $U_{\alpha i}$ is the MNS matrix.

Transition probability from
$\nu_{\alpha}$ to $\nu_{\beta}(\alpha\ne\beta)$
after having traveled a distance $L$ is given by
\begin{eqnarray}
P(\nu_{\alpha} \to \nu_{\beta})
=-4\sum^3_{i<j}{\rm Re}(J_{\alpha \beta}^{ij})
\sin^2 \phi_{ij}
-2JK
 \label{22},
\end{eqnarray}
where
\begin{eqnarray}
J_{\alpha \beta}^{ij}\equiv
U_{\alpha i}U_{\beta i}^* U_{\alpha j}^* U_{\beta j},
\quad 
\phi_{ij}\equiv \kappa \Delta m_{ij}^2L/E
=\kappa (m_i^2-m_j^2)L/E,
\end{eqnarray} 
with $\kappa=1/4$ in natural units or $\kappa=1.27$ 
in GeV per km and ${\rm eV}^2$ and 
\begin{eqnarray}
K\equiv 
4\sin \phi_{12}\sin \phi_{23}\sin \phi_{31}, 
\quad J\equiv {\rm Im}(J_{e\mu}^{12}).
\end{eqnarray}
$J$ is so called, Jarlscog factor \cite{Jarlscog},
${\rm Im}(J_{\alpha\beta}^{ij})$ obtained by the 
(anti-) cyclic permutation of ($e$, $\mu$) and ($1$,$2$) 
are all equal to ($-$)$J$.

\section{CP Asymmetry and Two Kinds of Triangles}
\hspace*{\parindent}
In this section, we would like to investigate CP asymmetry 
obtained from (\ref{22}) schematically.
We introduce two kinds of triangles. 
One is a unitarity triangle defined by 
$J_{\alpha\beta}^{ij}$ in (\ref{22}), 
which we call MNS triangle. 
The other is another new triangle defined by $\phi_{ij}$ 
also in (\ref{22}), 
which we call oscillation triangle.
We describe the definitions of the two kinds of triangles 
after we present our main results.
CP asymmetry is presented by using the sides of 
the two kinds of triangles as 
\begin{eqnarray}
A_{CP}&=&
\frac{P(\nu_{\alpha}\to\nu_{\beta})-
P(\bar{\nu}_{\alpha}\to\bar{\nu}_{\beta})}
{P(\nu_{\alpha}\to\nu_{\beta})
+P(\bar{\nu}_{\alpha}\to\bar{\nu}_{\beta})}
=\pm \frac{1}{\sqrt{1+D}} \label{5}
\end{eqnarray}
where $D$ is defined by
\begin{eqnarray}
4J^2K^2 D &\equiv& (|b|^2|n|^2-|c|^2|m|^2)(|a|^2|n|^2-|c|^2|l|^2)
\nonumber \\
&+&(|c|^2|l|^2-|a|^2|n|^2)(|b|^2|l|^2-|a|^2|m|^2) \nonumber \\
&+&(|a|^2|m|^2-|b|^2|l|^2)(|c|^2|m|^2-|b|^2|n|^2)
\label{16},
\end{eqnarray}
and $|a|$, $|b|$ and $|c|$ are the sides of an MNS triangle and 
$|l|$, $|m|$ and $|n|$ are an oscillation triangle as in Fig.1.

\begin{center}
\begin{minipage}{6cm}
\vspace*{0.5cm}
\includegraphics[scale=0.5]{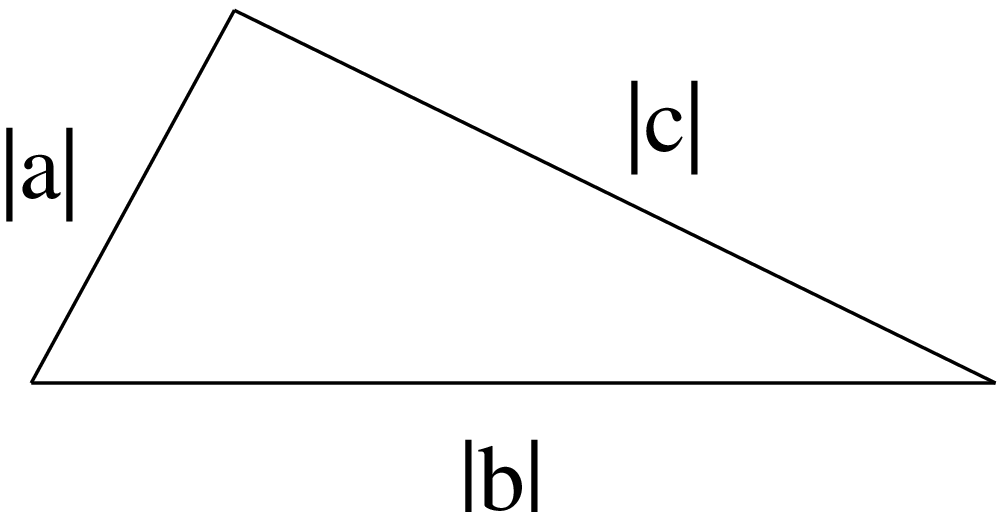}
\begin{flushleft}
Fig. 1(a) \quad an MNS triangle 
\end{flushleft}
\end{minipage}
\begin{minipage}{7cm}
\includegraphics[scale=0.5]{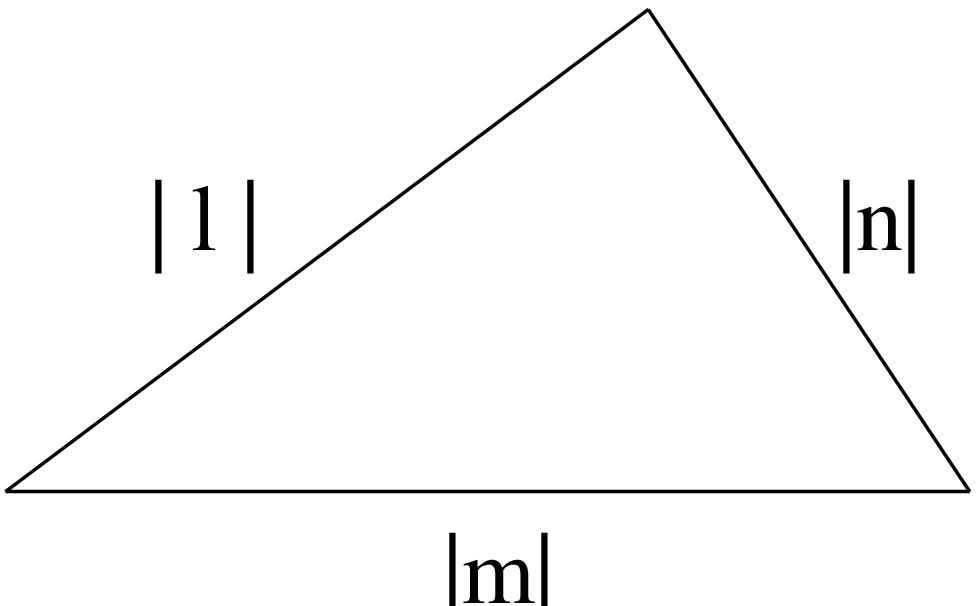}
\begin{flushleft}
Fig. 1(b) \quad an oscillation triangle
\end{flushleft}
\end{minipage}
\end{center}

The definition of the two kinds of triangles are as follows.
At first, we define MNS triangle 
by introducing complex numbers, $a$, $b$ and $c$ as
\begin{eqnarray}
a\equiv U_{\alpha 1}^* U_{\beta 1}, \qquad
b\equiv U_{\alpha 2}^* U_{\beta 2}, \qquad
c\equiv U_{\alpha 3}^* U_{\beta 3},
\end{eqnarray}
whose absolute values mean the length of the three sides 
of an MNS triangle because of unitarity condition, 
\begin{eqnarray}
U_{\alpha 1}^* U_{\beta 1}+U_{\alpha 2}^* U_{\beta 2}
+U_{\alpha 3}^* U_{\beta 3}=0.
\end{eqnarray}
$J_{\alpha\beta}^{ij}$ of (\ref{22}) are rewritten with 
the three sides 
of an MNS triangle from the relations such as 
\begin{eqnarray}
J_{\alpha\beta}^{12}=a^* b, \quad 
J_{\alpha\beta}^{23}=b^* c, \quad
J_{\alpha\beta}^{31}=c^* a \label{3}.
\end{eqnarray}

Next, we define oscillation triangle.
We choose one side (the base of the triangle whose length has no
physical meaning and is chosen arbitrary) and two base
angles, $\phi_{21}$ and $\phi_{32}$, as shown in Fig.2
to define the oscillation triangle.
Note that while the base is constant in baseline length, $L$,
both $\phi_{21}$ and $\phi_{32}$ are dependent on $L$
from the definition.
It should be emphasized that the position of point A
depends on $L$ and point A is located above or below
the base line of the triangle.
Oscillation triangles belong to either two classes like in Fig.2.
In Case I point A is located above the base like in Fig.2(a)
and in Case II point A is located below the base like in Fig.2(b).

\begin{center}
\begin{minipage}{6cm}
\vspace*{0.5cm}
\includegraphics[scale=0.5]{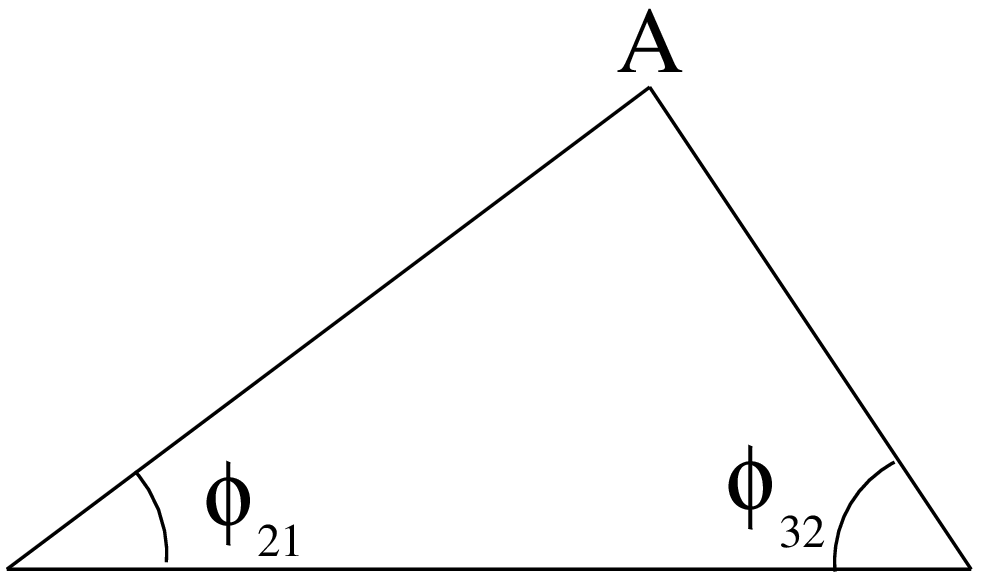}
\begin{flushleft}
Fig. 2(a) \quad Case I
\end{flushleft}
\end{minipage}
\begin{minipage}{6cm}
\includegraphics[scale=0.5]{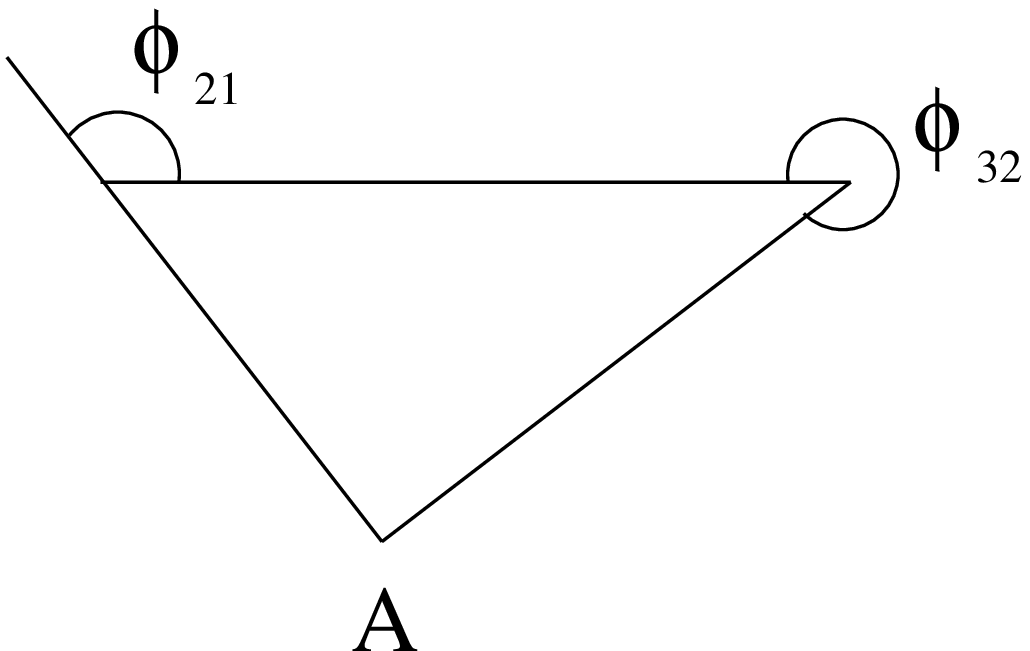}
\begin{flushleft}
Fig. 2(b) \quad Case II
\end{flushleft}
\end{minipage}
\end{center}

The interior angles of the oscillation triangles,
$\eta_1$, $\eta_2$ and $\eta_3$, are determined as
follows in each cases.

\begin{description}
\item[Case I] In Fig.2(a)
the interior angles are 
\begin{eqnarray}
\left\{
\begin{array}{l}
\eta_1= \phi_{32} -N\pi \label{27} \\
\eta_2= -\phi_{31} +(N+N^{\prime}+1)\pi \label{28} \\
\eta_3= \phi_{21} -N^{\prime}\pi \label{29}
\end{array}
\right.,
\end{eqnarray}
where $N$ and $N^{\prime}$ are integers which can be chosen
to make $\eta_1$ and $\eta_3$ between
$0$ and $\pi$.

\item[Case II] In Fig.2(b)
the interior angles are 
\begin{eqnarray}
\left\{
\begin{array}{l}
\eta_1= -\phi_{32} +N\pi \\
\eta_2= \phi_{31} +(1-N-N^{\prime})\pi  \\
\eta_3= -\phi_{21} +N^{\prime}\pi 
\end{array}
\right.,
\end{eqnarray}
where $N$ and $N^{\prime}$ are also integers which can be chosen
to make $\eta_1$ and $\eta_3$ between
$0$ and $\pi$.
\end{description}

As the transition probability of (\ref{22}) depends only 
on angles of the oscilletion triangle, 
the shape is important, 
on the other hand, the scale and the position  
in complex plane have nothing to do with 
the magnitude of transition probability.
We choose the radius of the circumcircle as $1$
and fix the oscillation triangles in complex plane 
as in Fig. 3 
making use of the arbitrariness of the scale and the position. 
The three sides, $l$, $m$ and $n$, are defined 
using complex numbers like 
\begin{eqnarray}
l=2\sin\eta_1(\cos\eta_3\mp i\sin\eta_3), \quad
m=2\sin\eta_2, \quad
n=2\sin\eta_3(-\cos\eta_1\pm i\sin\eta_1)
\end{eqnarray}
as the opposite side of the three angles, 
where the sign correspond to case I and case II.

\begin{center}
\begin{minipage}{6cm}
\vspace*{0.5cm}
\includegraphics[scale=0.5]{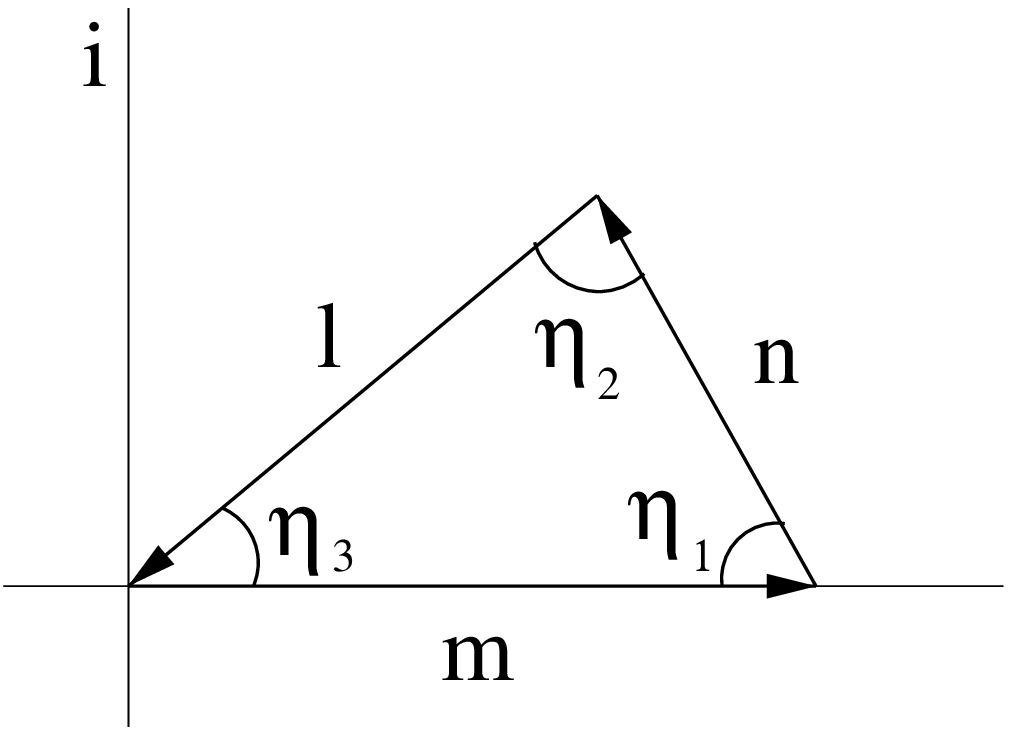}
\begin{flushleft}
Fig. 3(a) \quad Case I 
\end{flushleft}
\end{minipage}
\begin{minipage}{6cm}
\vspace*{0.5cm}
\includegraphics[scale=0.5]{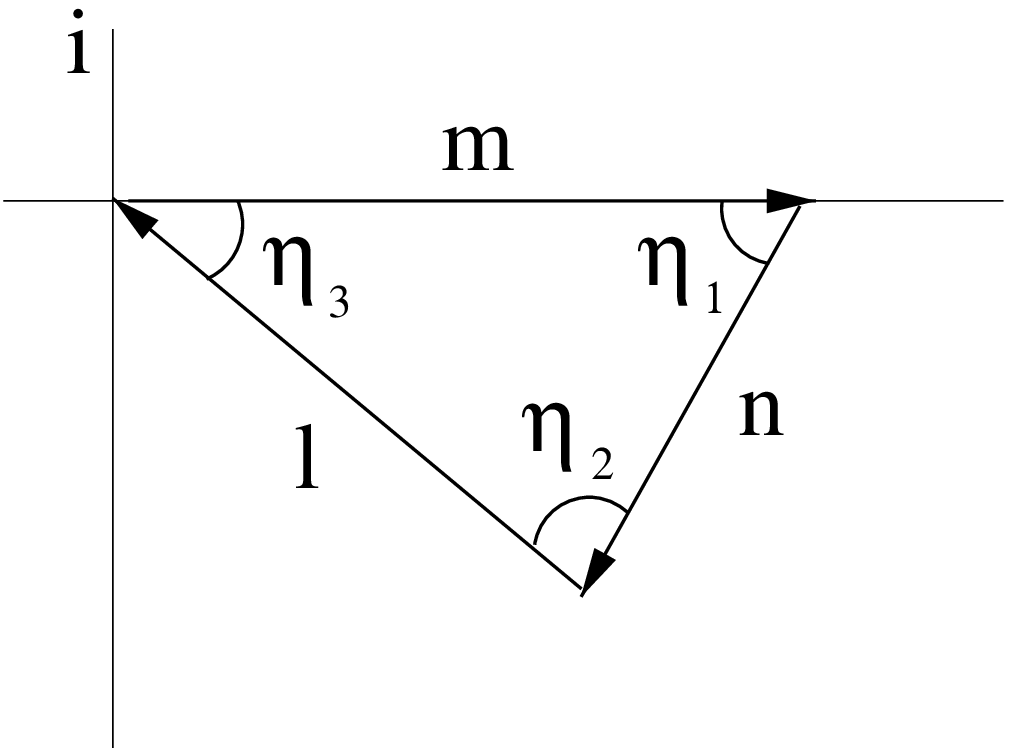}
\begin{flushleft}
Fig. 3(b) \quad Case II 
\end{flushleft}
\end{minipage}
\end{center}

We replace $\phi_{ij}$ of (\ref{22}) with the sides as 
\begin{eqnarray}
\sin^2 \phi_{23}=|l|^2/4, \quad \sin^2 \phi_{31}=|m|^2/4, 
\quad \sin^2 \phi_{12}=|n|^2/4 \label{4}.
\end{eqnarray}
Furthermore, 
the introduction of complex numbers as the three sides
makes possible to represent $K$ as the same form 
in both cases like 
\begin{equation}
K=\pm 4\sin \eta_1 \sin \eta_2 \sin \eta_3
={\rm Im}(l^* m)={\rm Im}(m^* n)={\rm Im}(n^* l),
\end{equation}
where plus and minus sign correspond to the Case I and Case II
respectively 
and the absolute value of $K$ means the twice of
the area of the oscillation triangle.

After the definition of the two kinds of triangles 
described above, we rewrite CP asymmetry, 
\begin{eqnarray}
A_{CP}=\frac{\Delta \CP}{\Delta P_{CP}}
=\frac{P(\nu_{\alpha} \to \nu_{\beta})
-P(\bar{\nu}_{\alpha} \to \bar{\nu}_{\beta})}
{P(\nu_{\alpha} \to \nu_{\beta})
+P(\bar{\nu}_{\alpha} \to \bar{\nu}_{\beta})},
\end{eqnarray}
using the sides of two kinds of triangles.
As $P(\bar{\nu}_{\alpha} \to \bar{\nu}_{\beta})$ 
are obtained by the replacement $U \to U^*$ in 
$P(\nu_{\alpha} \to \nu_{\beta})$,
CP-violating part and CP-conserving part are 
\begin{eqnarray}
 \Delta \CP&=&-4JK, \label{6} \\
\Delta P_{CP}&=&
-2\{{\rm Re}(a^*b)|n|^2+{\rm Re}(b^*c)|l|^2
+{\rm Re}(c^*a)|m|^2\},
\end{eqnarray}
where (\ref{3}) and (\ref{4}) are used.
Furthermore, CP-conserving part is rewritten as  
\begin{eqnarray}
(\Delta P_{CP}/2)^2&=&
4J^2K^2+(|b|^2|n|^2-|c|^2|m|^2)(|a|^2|n|^2-|c|^2|l|^2)
\nonumber \\
&&\quad \qquad 
+(|c|^2|l|^2-|a|^2|n|^2)(|b|^2|l|^2-|a|^2|m|^2) \nonumber \\
&&\quad \qquad 
+(|a|^2|m|^2-|b|^2|l|^2)(|c|^2|m|^2-|b|^2|n|^2)  
\label{7}
\end{eqnarray}
using the theorems about the sides and angles of 
triangles.
The important relations of (\ref{5}) and (\ref{16}) 
are obtained from (\ref{6}) and (\ref{7}), 
where $J \neq 0$ and $K \neq 0$ are used.
At the end of this section, we show the condition 
which gives the maximal CP asymmetry.
For the purpose, we consider another representation 
of $D$ like 
\begin{eqnarray}
 4J^2K^2 D&=& \frac{1}{|l|^2 |m|^2} 
\left( |c|^2 |l|^2 |m|^2 -{\rm Re}(m^* n)|b|^2 |l|^2
- {\rm Re}(n^* l)|a|^2 |m|^2\right)^2 \nonumber \\
&& +\frac{K^2}{4|l|^2 |m|^2}(|a|^2 |m|^2-|b|^2|l|^2)^2 
\geq 0
\label{42},
\end{eqnarray}
and from the equality condition of (\ref{42}) 
we have $100\%$ asymmetry,
\begin{equation}
A_{CP}=\pm 1,
\end{equation}
in the case that
the ratio of the sides of an MNS triangle and
an oscillation triangle is
\begin{equation}
|a|:|b|:|c|=|l|:|m|:|n| \label{17},
\end{equation}
namely the shapes of the two triangles
are the same.
Conversly, CP asymmetry is not $100\%$ if 
the two triangles are not the same shape,  
then the differences of the shapes determine the magnitude 
of CP asymmetry.

\section{Estimation of CP Asymmetry}
\hspace*{\parindent}
In this section, let us apply the schematic method 
obtained in the previous section 
for $\nu_e$-$\nu_{\mu}$ oscillation and estimate 
the magnitude of CP asymmetry.
In the following discussions, 
the representative set of parameters as
in Table.1 \cite{Fermilab} are used in 
LMA MSW and SMA MSW scenarios as examples.
Only upper bound, $\sin^2 2\theta_{13}<0.2$, is
given by CHOOZ experiment \cite{CHOOZ} about
$1$-$3$ mixing.
Later we consider how $A_{CP}$ changes
when $\sin \theta_{13}$ changes.
\begin{table}
\begin{center}
\begin{tabular}{ccccccc}
\hline \hline
Scenario & $\Delta m_{32}^2({\rm eV}^2)$ &
$\Delta m_{21}^2({\rm eV}^2)$ &
$\sin^2 2\theta_{23}$ & $\sin^2 2\theta_{12}$ &
$\sin^2 2\theta_{13}$ & $\delta$ \\
\hline
 & (atmos) & (solar) & (atmos) & (solar) & & \\
LMA MSW & $3.5 \times 10^{-3}$ &
$5 \times 10^{-5}$ & $1$ & $0.8$ & $0.04$ &
$\pi/2$ \\
SMA MSW & $3.5 \times 10^{-3}$ &
$1 \times 10^{-5}$ & $1$ & $0.01$ & $0.04$ &
$\pi/2$ \\
\hline \hline
\end{tabular}
\caption{reference set of mass square differences, 
mixing angles and CP phase}
\end{center}
\end{table}

There are two distinct mass differences
in three generations and they are set to
explain the atmospheric and the solar neutrino
deficit.
The form of the MNS matrices in these two scenarios are
\begin{equation}
U({\rm LMA})= \left(
\begin{array}{ccc}
0.846 & 0.523 & -0.101i \\
-0.372-0.060i & 0.602-0.037i & 0.704 \\
0.372-0.060i & -0.602-0.037i & 0.704
\end{array}
\right)
\end{equation}
for LMA MSW scenario
and
\begin{equation}
U({\rm SMA})= \left(
\begin{array}{ccc}
0.994 & 0.050 & -0.101i \\
-0.035-0.071i & 0.706-0.004i & 0.704 \\
0.035-0.071i & -0.706-0.004i & 0.704
\end{array}
\right)
\end{equation}
for SMA MSW scenario.
In each cases, the MNS triangle corresponding to
$\nu_e$-$\nu_{\mu}$ oscillation is shown in
Fig. 4(a) and Fig. 4(b).
The three sides of the MNS triangle are 
\begin{eqnarray}
(|a|,|b|,|c|)=\left\{
\begin{array}{cc}
(0.32, 0.32, 0.071) & {\rm for \, LMA \, MSW} \\
(0.079, 0.035, 0.071) & {\rm for \, SMA \, MSW} 
\end{array}
\right.
\end{eqnarray}
and three angles are 
\begin{eqnarray}
(\psi_1,\psi_2,\psi_3)=\left\{
\begin{array}{cc}
(1.46, 1.46, 0.22) & {\rm for \, LMA \, MSW} \\
(1.56, 0.46, 1.12) & {\rm for \, SMA \, MSW} 
\end{array}
\right..
\end{eqnarray}

\begin{center}
\begin{minipage}{7cm}
\vspace*{0.5cm}
\includegraphics[scale=0.8]{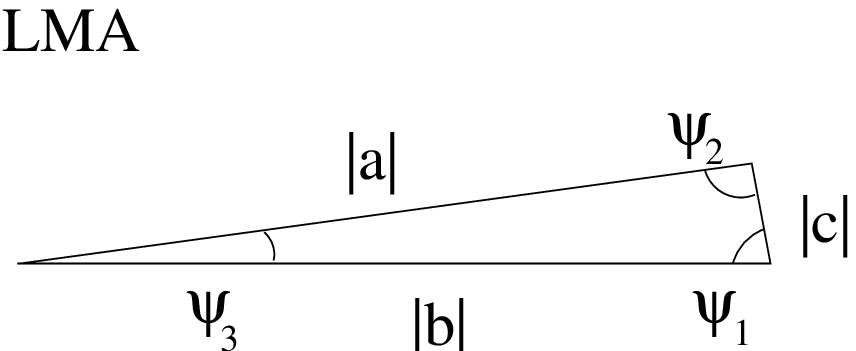}
\begin{flushleft}
Fig. 4(a) \quad MNS triangle for LMA MSW scenario
\end{flushleft}
\end{minipage}
\hspace*{1cm}
\begin{minipage}{5cm}
\vspace*{0.5cm}
\includegraphics[scale=0.8]{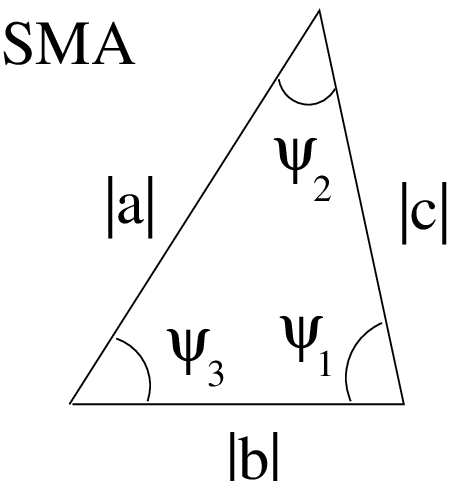}
\begin{flushleft}
Fig. 4(b) \quad MNS triangle for SMA MSW scenario
\end{flushleft}
\end{minipage}
\end{center}

At first, let us show that 
the smallest $L/E$ which gives the almost
$100\%$ asymmetry is estimated as 
\begin{eqnarray}
\frac{L({\rm km})}{E({\rm GeV})}=\left\{
\begin{array}{cc}
3200 & {\rm for \, LMA \, MSW} \\
88000 & {\rm for \, SMA \, MSW}
\end{array}
\right. \label{2},
\end{eqnarray}
with the schematic method.
These values are derived as following two steps.
At first, choose $L/E$ such that 
$\eta_3=\phi_{21}=\psi_3$, that is   
\begin{eqnarray}
\phi_{21}=\frac{1.27\Delta m_{21}^2 L}{E}=
\left\{
\begin{array}{cc}
0.22 & {\rm for \, LMA \, MSW} \\
1.12 & {\rm for \, SMA \, MSW}
\end{array}
\right. \label{1}.
\end{eqnarray}
Second, change the value of $L/E$ obtained from (\ref{1}) a little
so that $\eta_1=\phi_{32}-N\pi=\psi_1$, that is 
\begin{eqnarray}
\phi_{32}=\frac{1.27\Delta m_{32}^2 L}{E}=
\left\{
\begin{array}{cc}
1.46+N_L\pi 
& {\rm for \, LMA \, MSW} \\
1.56+N_S\pi 
& {\rm for \, SMA \, MSW}
\end{array}
\right.,
\end{eqnarray}
where $N_L=4$ and $N_S=124$ for the $L/E$ determined by 
(\ref{1}).
There is little change for the angle $\eta_3$ set at first 
by the second operation 
because  $\phi_{32}$ and $\phi_{31}$ change rapidly compared
to $\phi_{21}$.
Thus, we can easily estimate $L/E$ which gives almost 
$100\%$ asymmetry as (\ref{2}) by choosing 
the oscillation triangle to be the almost
same shape as the MNS triangle in Fig. 4.

Next, let us roughly estimate whether $L/E$ of (\ref{2}) 
can be realized in future experiments 
taking the two kinds of experiments as examples.
First, in the experiments of neutrino factory 
\cite{neutrino factory},
suppose that the energy of neutrino beam is 
$5\, {\rm GeV}$
and the baseline length is about $7400$ km from 
Fermilab to Gran Sasso. 
Second, in the experiments of PRISM \cite{PRISM},
suppose that the energy of the neutrino beam is 
$100\,{\rm MeV}$
and the baseline length is $L\sim 250\, {\rm km}$
from KEK to Super-Kamiokande.
The value of $L({\rm km})/E({\rm GeV})$ realized 
in the two kinds of future experiments are 
\begin{eqnarray}
\frac{L({\rm km})}{E({\rm GeV})}
\sim 
\left\{
\begin{array}{cc}
1500 & {\rm for \, neutrino \,factory} \\
2500 & {\rm for \, PRISM}
\end{array}
\right..
\end{eqnarray}
These values of $L({\rm km})/E({\rm GeV})$ are close to  
the value of (\ref{2}) for LMA MSW scenario. 
Therefore, it is expected to obtain rather large 
CP asymmetry.
Actually at the value of $L({\rm km})/E({\rm GeV})\sim 2500$
we obtain 
\begin{eqnarray}
A_{CP}\sim 
\left\{
\begin{array}{cc}
90\% & {\rm for \, LMA \, MSW} \\
3\% & {\rm for \, SMA \, MSW}
\end{array}
\right.,
\end{eqnarray}
where 
\begin{eqnarray}
(|l|,|m|,|n|)=\left\{
\begin{array}{cc}
(1.99, 1.92, 0.32) & {\rm for \, LMA \, MSW} \\
(1.99, 1.98, 0.063) & {\rm for \, SMA \, MSW}. 
\end{array}
\right.,
\end{eqnarray}
derived from the definition of $\phi_{ij}$ and (\ref{4}) 
are used.

Finally, let us consider how the magnitude of $A_{CP}$ 
is changed by $\sin \theta_{13}$.
If $\sin \theta_{13}$ is smaller than the value we choose, 
$|c|$ and the angle $\gamma$ corresponding to $|c|$ of 
the MNS triangle is also small.
Eq.(\ref{1}) leads to small $L/E$ which gives 
almost $100\%$ asymmetry according to the smallness of $\gamma$.
Therefore, rather large $A_{CP}$ may be realized
not only in LMA MSW scenario but also in SMA MSW
scenario if $\sin \theta_{13}$ is small.
However, as small $\gamma$ means the squashed MNS triangle, 
$J$ and therefore $\Delta \CP$ become also small. 
We need the beam which has high intensity enough to compensate 
the smallness of $\Delta \CP$ to observe $A_{CP}$.

\section{Summary and Discussions}

\hspace*{\parindent}

In the framework of three generations
we have mainly investigated CP asymmetry in neutrino oscillations.
The main results of this paper are following;

\begin{enumerate}
\renewcommand{\labelenumi}{(\roman{enumi})}

\item  We propose the scematical method to estimate
the magnitude of $A_{CP}$.
$A_{CP}$ is calculated from the length of three sides
of an MNS triangle and an oscillation triangle.
$A_{CP}$ takes the maximal value in the case that
the shapes of the two triangles are the same.

\item  In vacuum $\nu_e$-$\nu_{\mu}$ oscillation,
we obtain almost $100\%$ asymmetry
in $L({\rm km})/E({\rm GeV})\sim 3200$ for LMA MSW scenario and
in $L({\rm km})/E({\rm GeV})\sim 88000$ for SMA MSW scenario.

\item  In PRISM experiment to be expected to realize
near future,
we obtain about $A_{CP}\sim 90\%$ for LMA MSW
scenario and $A_{CP}\sim 3\%$ for SMA MSW
scenario with $\sin^2 2\theta_{13}\simeq 0.04$.

\end{enumerate}

Finally we note the prospect in future.
In order to determine the magnitude of CP asymmetry 
we need to estimate other parameters of the MNS matrix 
as precise as possible.
Furthermore, 
the mass differences and mixing angles are changed
by the earth matter effect in practical
long baseline neutrino experiments.
We would like to investigate how large the earth matter effect is
in CP violation and in what region our results 
are ineffective in the paper to be prepared.
We will also study $A_{CP}$ averaging in 
the neutrino energy and the baseline length 
in order to estimate CP violating effects 
in practical experiments.
If the limitation of the energy resolution 
is improved, the CP violating signals become more sharp.
We should consider the better method to obtain 
the CP violating signal as sharp as possible.
In addition, we investigate other MNS triangles
in the next paper.

\vspace{20pt}
\noindent
{\Large {\bf Acknowledgement}}

\noindent
The authors would like to thank Prof. A. I. Sanda for
useful discussions and fruitful comments.
We would like to thank Prof. Y. Sugiyama for careful reading
of manuscript.

\end{document}